\documentclass[]{revtex4}
\usepackage[dvips]{graphicx}

\begin{document}

\title{
Novel ordering of the pyrochlore Heisenberg antiferromagnet with the ferromagnetic next-nearest-neighbor interaction
}

\author{Daisuke Tsuneishi, Masayuki Ioki and Hikaru Kawamura}

\address{Faculty of Science, Osaka University, Toyonaka 560-0043, Japan}
\begin{abstract}
The ordering property of the classical pyrochlore Heisenberg antiferromagnet with the  ferromagnetic next-nearest-neighbor interaction is investigated by means of a Monte Carlo simulation. The model is found to exhibit a first-order transition at a finite temperature into a peculiar ordered state. While the spin structure factor, {\it i.e.\/}, the thermal average of the squared Fourier amplitude of the spin, exhibits a finite long-range order characterized by the commensurate spin order of the period four, the thermal average of the spin itself almost vanishes. It means that, although the amplitude of the spin Fourier component is long-range ordered, the associated phase degree of freedom remains to be fluctuating.
\end{abstract}

\maketitle

\section{Introduction}

Recently, there arises a considerable interest in the ordering of pyrochlore magnets. The pyrochlore lattice consists of corner-sharing tetrahedra, and is generally  regarded as a typical realization of geometrically frustrated lattices. In particular, the classical Heisenberg antiferromagnet on the pyrochlore lattice is known to exhibit no magnetic long-range order of any kind even at zero temperature, if the interaction is restricted only to nearest neighbors \cite{Reimers2,Moessner,Isakov,Heneley}. This is due to the extremely high degeneracy of the ground state induced by severe geometrical frustration. Such a high degeneracy is realized on the delicate balance among frustrated interactions, and might be lifted by the weak perturbation which inevitably exists in real magnets, {\it e.g.\/}, the further neighbor interactions, the dipolar interaction, the magnetic anisotropy, the quenched randomness, or the lattice distortion {\it etc.} The lifting of the degeneracy would eventually lead to a phase transition at a low but nonzero temperature, possibly of an exotic type peculiar to geometrically frustrated magnets. Thus, it is important to clarify the role of various weak perturbations on the ordering of the nearest-neighbor Heisenberg model.

Some time ago, Reimers performed a pioneering study of the ordering of the classical Heisenberg pyrochlore antiferromagnet with the antiferromagnetic first-neighbor interaction $J$ and the further-neighbor interactions (up to the fourth-neighbors), $J_2\sim J_4$, within a mean-field approximation \cite{Reimers1}. He constructed a possible phase digram as a function of the interaction parameters $J_2/|J|\sim J_4/|J|$, revealing a variety of ordered phases. 

In the present paper, we investigate the ordering property of the classical pyrochlore Heisenberg antiferromagnet with the {\it ferromagnetic next-nearest-neighbor interaction\/} $J_2>0$ by means of a Monte Carlo simulation \cite{JSPS}. In fact, we also studied the same model with the {\it antiferromagnetic\/} next-nearest-neighbor interaction $J_2<0$, to find that the system exhibited a first-order transition into the $q=0$  ordered phase with the collinear up-up-down-down spin structure \cite{Reimers2,JSPS}. There, the ordered state is conventional in the sense that the long-range magnetic order sets in. In the case of the ferromagnetic next-nearest-neighbor interaction $J_2>0$, by contrast, while the system also exhibited a first-order transition at a finite temperature into the ordered state, this ordered state turns out to be quite peculiar, as we shall see below.

\section{Model and physical quantities}

The model we consider is the classical Heisenberg model on a pyrochlore lattice with the antiferromagnetic first-neighbor interaction $J<0$ and the ferromagnetic second-neighbor interaction $J_2>0$. The Hamiltonian is given by
\begin{equation}
   {\cal H} = - J \sum_{<i,j>} \vec{S}_i \cdot \vec{S}_j 
       - J_2 \sum_{<k,l>} \vec{S}_k \cdot \vec{S}_l, 
\end{equation}
where $\vec S_i$ is a three-component classical Heisenberg spin variable at the $i$-th site, $\vec S_i=(S_i^x,S_i^y,S_i^z)$ with $|\vec S_i|=1$, and the sum is taken over all first-neighbor and second-neighbor pairs on the lattice.

We study the ordering properties of the model by Monte Carlo simulations. The strength of the second-neighbor interaction is set to $J_2=0.1|J|$. 
(In fact, we have also studied other values of $J_2$, to find that the result is insensitive to the $J_2$-value, at least qualitatively.) 
At lower temperatures, a very slow dynamics, presumably originated from the intrinsic frustration effect, arises. To circumvent the resulting thermalization problem, we adopt in our simulation a temperature-exchange method 
%\cite{Huku} 
combined with the standard heat-bath updating.

In a cubic unit cell of the pyrochlore lattice, there are 16 sites (spins). We measure the linear size of our system $L$ in units of the cubic unit cell, {\it i.e.\/}, our system contains $N=16L^3$ spins in total. Periodic boundary conditions are imposed in all directions.

On a pyrochlore lattice, complex spin order could arise due to the severe frustration effect. Accordingly, in order to detect the possible complex spin order, we calculate the spin freezing parameter, a quantity familiar in the study of spin glasses, defined by
\begin{equation}
   q_s^{(2)} = \sum_{\mu,\nu} \langle q_{\mu \nu}^2 \rangle,~~~(\mu,\nu = x,y,z), \ \ \ \    
q_{\mu\nu} = \frac{1}{N} \sum_i S^{(a)}_{i\mu} S^{(b)}_{i\nu},
\end{equation}
where $\langle \cdots \rangle$ is the thermal average, and the upper suffixes (a) and (b) denote the two copies (replicas) of the system. In order to compute  $q_s^{(2)}$, we simulate two independent replicas of the system, (a) and (b), by using different initial conditions and different random-number sequences. In equilibrium,  $q_s^{(2)}$ is identically given by
\begin{equation}
   q_s^{(2)} = \frac{1}{N^2} \sum_{i,j} \langle \vec S_i \cdot \vec S_j \rangle ^2. 
\end{equation}
As is evident from Eq.(3), the spin freezing parameter $q_s^{(2)}$ becomes unity for any ordered state where the mutual direction of spins is completely  frozen, whatever type of ordered state it may be. If the mutual direction of spin remains fluctuating, on the other hand, $q_s^{(2)}$ tends to vanish. Note that $q_s^{(2)}$ is invariant under continuous symmetry operations of the Heisenberg Hamiltonian, {\it i.e.\/}, under global spin rotations {\it made independently of the two replicas (a) and (b)\/}. Such an invariance of  $q_s^{(2)}$ under global spin rotations is practically important, since, in finite-size simulations of the isotropic Heisenberg model, global spin rotations easily occur in the course of simulation. Concerning the discrete symmetry of the Hamiltonian, $q_s^{(2)}$ is invariant under global spin reflections, but is not necessarily so under discrete symmetry operations associated with the underlying lattice symmetry, {\it e.g.\/}, lattice translations, which can not be adsorbed into spin rotations nor reflections. 

In order to monitor the type of spatial spin order, we also calculate the spin structure factor $F(\vec{q})$, {\it i.e.\/}, the thermal average of the squared Fourier amplitude defined by
\begin{equation}
   F(\vec{q}) = \frac{1}{N^2} \bigl\langle \bigl| \sum_j \vec{S}_j~e^{i\vec{q}
   \cdot \vec{r}_i} \bigr|^2 \bigr\rangle
\end{equation}
Other standard quantities, {\it e.g.\/}, the energy, the specific heat and the susceptibility, {\it etc\/} are also calculated.

\section{Monte Carlo results}

In  Fig.1, we show the temperature and size dependence of the energy per spin. As can be seen from the figure, an almost discontinuous jump characteristic of a first-order transition is observed at $T=T_N\simeq 0.094J$, which becomes increasingly eminent for larger lattices. The calculated energy histogram (not shown here) exhibits a pronounced double peak structure at around $T=T_N$, demonstrating that the transition is certainly first order.

\begin{figure}[h]
\begin{center}
\includegraphics[scale=0.38]{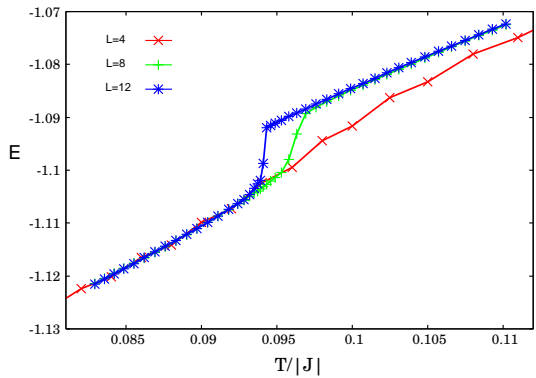}
\end{center}
\caption{
The temperature and size dependence of the energy per spin.
}
\end{figure}

\begin{figure}[h]
\begin{center}
\includegraphics[scale=0.68]{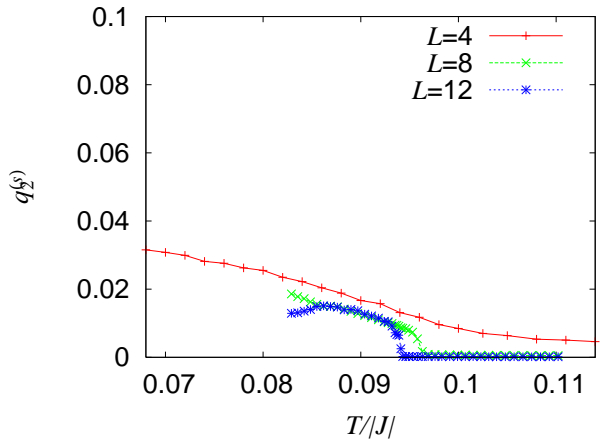}
\end{center}
\caption{
The temperature and size dependence of the spin freezing parameter $q_s^{(2)}$.
}
\end{figure}

In Fig.2, we show the temperature and size dependence of the freezing parameter $q_s^{(2)}$. As can be seen from the figure, the growth of $q_s^{(2)}$ is rather week; $q_s^{(2)}$ takes small values of order $10^{-2}$ even below the first-order transition point. Indeed, even when we extrapolate the observed temperature dependence of $q_s^{(2)}$ down to $T=0$, it yields only 10\% of its full value or less.  Whether $q_s^{(2)}$ in the ordered state entirely vanishes or keeps a small nonzero value in the thermodynamic limit $L\rightarrow \infty$, is not entire clear. While the data of larger lattices $L=8$ and 12 suggest that a small nonzero value persists even in the thermodynamic limit, further careful study is necessary to clarify this point. Anyway, the spin remains largely fluctuating even in the ordered state.

We also calculate the spin structure factor $F(\vec q)$ at wavevectors $\vec q = \frac{2\pi}{a}(h,h,l)$  at a temperature $T=0.084|J|$ below $T_N$ ($a$ being the spacing of the cubic unit cell), which is shown in Fig.3 in the ($l$,$h$) plane. Main peaks are observed at $(\pm 5/4, \pm 5/4, 0)$ and at $(\pm 3/4, \pm 3/4, 2)$, with sub-main peaks at $(\pm 3/4, \pm 3/4, 0)$ and at $(\pm 5/4, \pm 5/4, 2)$. Note that, in the whole $\vec q$-space, there are twelve independent points at each $(5/4, 5/4, 0)$, $(3/4, 3/4, 2)$, $(3/4, 3/4, 0)$ and $(5/4, 5/4, 2)$, respectively. These peaks turn out to be sharp, with very little amplitude left at other positions of $\vec q$. The structure factor along the $(h,h,0)$ direction is shown in Fig.4 for various lattice sizes $L$. These data clearly indicate that, when probed via the structure factor, the system exhibits an almost complete commensurate magnetic long-range order of the period four. We note that such a period-four commensurate spin order is at odd with the one predicted in Ref.\cite{Reimers1} based on the mean-field analysis.

Our observation for the spin structure factor is in apparent contradiction to the observed near-vanishing of the spin freezing parameter $q_s^{(2)}$. It means that, although the spin itself remains largely fluctuating even in the ordered state, the system is almost fully ordered when probed via the spin structure factor. Since the structure factor measures only the amplitude of the spin Fourier component irrespective of the state of its phase degree of freedom, the ordered state is a peculiar one where the Fourier amplitude of the spin exhibits a long-range order while the phase of the spin Fourier component remains fluctuating. 
%To the authors' knowledge, this is the first time that such a peculiar ordered state is identified in model calculation. 

We also examined the snapshot of spin patterns realized at lower temperature, to find that the period-four periodicity of the ordered state is hardly visible in the spin pattern itself consistently with our observation of $q_s^{(2)}$, which, however, might partly be due to the finite-temperature effect.

\begin{figure}[h]
\begin{center}
\includegraphics[scale=1.4]{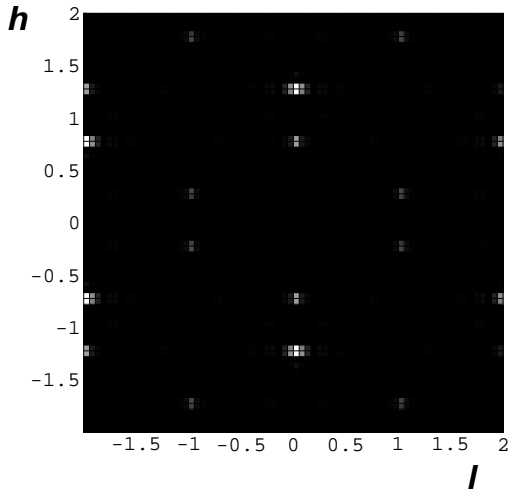}
\end{center}
\caption{
The spin structure factor $F(\vec{q})$ at wavevectors 
$\vec q = \frac{2\pi}{a}(h,h,l)$  shown in the ($l,h$) plane. 
The temperature is $T=0.084|J|$ below $T_N$.
}
\end{figure}

\begin{figure}[h]
\begin{center}
\includegraphics[scale=0.43]{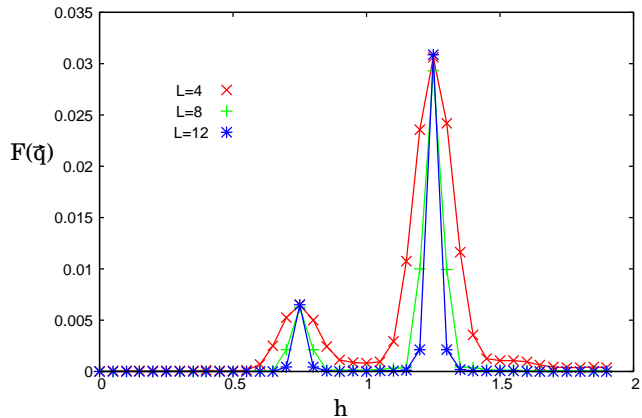}
\end{center}
\caption{
The $h$-dependence of the spin structure factor $F(\vec{q})$ along the $(h,h,0)$ direction for various lattice sizes $L$. The temperature is $T=0.084|J|$. 
}
\end{figure}

One may wonder if the observed near-vanishing of $q_s^{(2)}$ might still be an artifact due to finite-size effects. For instance, although $q_s^{(2)}$ is invariant under both global spin rotations and reflections as mentioned above, possible global sliding motion of the ordered state associated with lattice translations might tend $q_s^{(2)}$ to vanish. In order to examine such a possibility, we also calculate $q_s^{(2)}$ by applying an appropriate symmetry-breaking field of intensity $H_{sb}$ to the system,  $- H_{sb} \sum_i \vec {S_i}\cdot \vec{h}^{sb }_i$, 
%
%\begin{equation}
%   {\cal H}^\prime = {\cal H} - H_{sb} \sum_i \vec {S_i}\cdot \vec{h}^{sb}_i,
%\end{equation}
%
where the direction of the symmetry-breaking field $\vec{h}^{sb}_i$ (with $|\vec{h}^{sb}_i|=1$) is chosen to be the spin direction in the lowest-temperature configuration generated in separate simulation with $H_{sb}=0$. If the observed near-vanishing of $q_s^{(2)}$ is a finite-size effect associated with the global sliding of the ordered state, the application of the symmetry-breaking field would quickly pin down such a global sliding, leading to the appearance of larger $q_s^{(2)}$ values of order unity when the temperature is lowered across the transition temperature of the $H_{sb}=0$ model. Our calculation with $H_{sb}=0.1|J_2|$, however, indicates that the induced $q_s^{(2)}$ still remains small, without any rapid growth being observed around the transition temperature of the $H_{sb}=0$ model: The $q_s^{(2)}$ value extrapolated to $T=0$ is about $0.3$ for $L=4$.  These observations might suggest that the observed near vanishing of $q_s^{(2)}$ is a bulk phenomenon, not merely a finite-size effect due to the global sliding of the ordered state.

\section{Summary}

The ordering property of the classical pyrochlore Heisenberg antiferromagnet with the ferromagnetic next-nearest-neighbor interaction is investigated by means of a Monte Carlo simulation. The model turns out to exhibit a first-order transition into a peculiar ordered phase where the spin structure factor, {\it i.e.\/}, the thermal average of the squared Fourier amplitude, exhibits a finite long-range order characterized by the commensurate spin order of the period four, while the thermal average of the spin itself almost vanishes.

We finally mention that an exotic ordered state as revealed here might have important implications to experiments. In such an exotic ordered state, while the conventional Bragg peak would be observed in neutron-scattering measurements, no magnetic long-range order would be observed in NMR or $\mu$SR measurements. Experimental quest for such an exotic ordered state in real pyrochlore materials would be of utmost interest. 

The author is thankful to Dr. M.J.P. Gingras for useful discussion.

%Appearance of such an anomalous ordered state is in sharp contrast to the corresponding case of the antiferromagnetic neaxt-nearest neighbor interaction. 

\end{document}